\documentclass[twocolumn,showpacs,aps,prd]{revtex4-1}

\usepackage{graphicx}
\usepackage{mdwlist}
\usepackage{dcolumn}
\usepackage{multirow}
\usepackage{amsmath}
\usepackage{epsfig}
\usepackage{subfigure}

\def\figurebox#1#2#3{%
    \def\arg{#3}%
    \ifx\arg\empty
    {\hfill\vbox{\hsize#2\hrule\hbox to #2{\vrule\hfill\vbox to #1{\hsize#2\vfill}\vrule}\hrule}\hfill}%
    \else
    {\hfill\epsfbox{#3}\hfill}%
    \fi}

\begin{document}

\title{
{\large \bf
The Two  Subcultures: The teaching of theory and Physics' place in the college curriculum
}
}

\newcommand*{\EMAIL}{mmccracken@washjeff.edu}
\author{M.E.~McCracken}
\altaffiliation[E-mail: ]{\EMAIL}
\affiliation{Washington \& Jefferson College, Washington, Pennsylvania 15301}


\date{\today}
\begin{abstract}
During recent collaboration with colleagues
to revise our institution's general-education curriculum, I encountered many perceptions of what we mean by {\em the Natural Sciences}.
I was surprised to find that perceptions of scientific pedagogy varied significantly among the scientific disciplines, especially concerning issues of philosophy of science and epistemology, manifested in the approaches to teaching theoretical concepts and their development.
These realizations suggest that Physics occupies a singular role in college curricula, introducing students, even at the introductory level, to the acquisition of knowledge by theoretical means and the assessment of theory based on experimental evidence.
\end{abstract}
\maketitle

I am a physics professor at a small liberal arts college in the mid-Atlantic states, where in the last several years I have worked with colleagues from other departments to review and revise the college's general-education curriculum.
This has been a great experience; I've learned much from my colleagues, and I believe that my appreciation for liberal education and my ability to be an effective academic advisor have increased dramatically.
I've benefited from being a part of many fascinating discussions of pedagogical approach involving passionate educators.
I think that my future physics classes will be enriched by techniques that I'll be borrowing 
from colleagues across campus, including those as seemingly remote as 
the Arts and Humanities. 

However, with the review process in its final stages, I've encountered perceptions of the Natural Sciences that have left me baffled and concerned about how we present the Sciences to our students and the public.
These have not come, as one might assume, from faculty in non-science disciplines, but from faculty {\em in} the Natural Sciences.
This situation has confused me about the way I've approached my teaching and my scholarship in physics.
I've found that colleagues from other disciplines have opinions on this, too, and the comments that I make below should be accessible to anyone.

To set the stage for my argument (and to demonstrate what is at stake), I want to begin with a story with which the reader is likely familiar, one which I consider to be the quintessence of theoretical inquiry.  

\subsection{Einstein and the development of General Relativity}

In 1907, Albert Einstein was thinking about gravity.\footnote{The history here is adapted from several sources. References \cite{zee,ferris} provide excellent and entertaining treatments.}
Two years earlier, Einstein had what is now referred to as his {\em annus mirabilis}. The 1905 publication of his ground-breaking articles on special relativity, Brownian motion, and the photoelectric effect built the foundation for Einstein's place as the most prolific theoretical physicist of the 20$^{\textrm{th}}$ century.
But in 1907, midway on his trajectory from the outer fringe to the center of the physics community, Einstein was troubled by gravity.

At the time, the working theory of gravitation was that of Isaac Newton, dating back to at least 1687.
Newtonian gravity can be captured in the form of a single law: the attractive gravitational force between two objects is proportional to their masses and inversely proportional to the square of the distance between them.
Symbolically, we write that
\begin{equation}\label{e:grav}
F_{g} = G\frac{m_1 m_2}{r^2},
\end{equation}
where $F_g$ is the attractive force between two objects with masses $m_1$ and $m_2$ separated by distance $r$, and $G$ is a proportionality constant that quantifies the strength of the interaction.
($G$ has the same value for all pairs of objects.)

This law has many consequences, some of which are broad and conceptual.
For example, this law suggests that {\em any} two objects with mass gravitationally attract, {\em i.e.}, that gravitational attraction is a {\em universal} phenomenon.  
This is a revolutionary idea!
If you are reading this essay on a computer there is an attractive gravitational force between it and you, but it's roughly twenty orders of magnitude ({\em i.e.}, a hundred-billion-billion times) smaller than the gravitational force that the Earth exerts on you.
Newton's gravitational law unified seemingly unrelated events such as the falling of an apple or the streaking of a comet across the night sky, revealing that they are manifestations of the same gravitational interaction.

Some of the consequences of Newton's theory are more specific.  
The details of eq.~\ref{e:grav} are quite interesting, but it is the {\em existence} of eq.~\ref{e:grav} that is more important to our point.
The mathematical nature of the universal gravitation law allows us to make quantitative predictions and compare these to observations.
Newtonian gravity has had immense successes when used in this way.
For many phenomena, on scales from those of solar systems  to bench-top laboratory experiments (see {\em e.g.}~\cite{Adelberger2009102}), Newton's theory of gravitation can predict results that are consistent with experimental measurements.
For many modern problems such as planning interplanetary satellite flight paths \footnote{A particularly stunning example is the trajectory of the Cassini satellite.  See \cite{Wolf19951611}.} application of Newton's 300-year-old law is sufficient!

However, Newtonian gravity does present some glaring problems of both the broad and specific types described above.
One of the more compelling specifics, known since the 1860's, was the precession of the orbit of Mercury: as Mercury orbits the Sun, its elliptical trajectory shifts in a way that can not be accounted for by Newtonian theory.
By the late 1800's there were possible explanations that could reconcile this observation within Newton's framework (e.g., other hidden planets affecting Mercury's orbit), none of these panned out.

A perhaps more fundamentally troubling aspect of Newton's theory is the consequence that the acceleration of an object in a gravitational field ({\em i.e.}, the rate at which its velocity changes under the influence of gravity) does {\em not} depend on the object's mass, even though the gravitational force does.
(If you've taken a physics class, this is why we talk of {\em the} acceleration due to gravity (9.8~m/s$^{2}$) for objects near the Earth's surface.)
This feature of gravity is one of the first things that we teach students in an introductory physics course, but it was (before Einstein) a deep mystery and a suspicious coincidence.

In 1907, while still working as a patent clerk, Einstein had what he later referred to as ``the happiest thought of [his] life.''  
The Thought is an incredibly simple one: a person undergoing free-fall ({\em i.e.}, falling under the influence of {\em only} gravity) does not feel her own weight.
Einstein imagined a person falling off of a roof, but I encourage you to imagine an executive taking the elevator to her office when the elevator's support cables suddenly break.
If, out of surprise, she releases her briefcase as the elevator is falling she will observe that it does not leave the position in which she held it.
An external observer, of course, would say that both the executive and her briefcase were falling with the same acceleration.
Einstein's insight was to realize that to the person in this elevator, it would appear that no gravitational force acted on her or the briefcase.
The broader consequence of this thought is that there is no local test that the falling person can perform to distinguish between falling freely under the influence of gravity or floating in the absence of gravitational forces.
The same is true to the astronauts aboard the ISS; they appear weightless, even though they are constantly falling in a circular orbit around the Earth.

Fueled by The Thought, Einstein spent the next eight years developing the {\em general theory of relativity} (GR), a wedding of his earlier special relativistic theory and gravitation.
Einstein's development of GR is a paragon of what physicists mean when they refer to ``theoretical physics.'' 
His program was based on a conceptual insight, which was then expanded into a rich framework of mathematical and logical tools.
This work was motivated as much by mathematical consistency, necessity, and beauty as by physical observations.
Einstein pulled insights from areas of mathematics previously thought to be purely logical constructs to develop a new language for gravitation.
Einstein's work presented what appears to be {\em the} language of gravity at large scales.

In the same way as Newton's theory, GR gives us both broad, profound principles and specific predictions.
It makes definite, quantitative predictions that can be compared to experimental observations.
GR has predicted many novel phenomena (black holes, gravitational lensing, gravitational time dilation, {\em et c.}), some of which have already been observed and some of which we still seek.
In this way, GR is {\em testable}; there are definite tests that we can perform, which, if theoretical and experimental results disagreed, would indicate that GR is {\em not} the complete theory of gravity.
To date, there are some reasons to expect that GR is an incomplete description of gravity, but there are few experimental observations that GR cannot explain.
Despite its beauty, power, and success, I doubt that any physicist would claim that General Relativity is the end of the story.

One more note, just for fun.
Einstein was famously sardonic when commenting on experiment.
In 1919, a huge expedition was undertaken to make measurements of a phenomenon that was predicted by GR, but was outside the realm of Newtonian gravity: the deflection of starlight by the Sun during the May 29 solar eclipse.
Though GR passed this first empirical test with flying colors, Einstein was seemingly cool.
When asked how he would have reacted had the experiment shown no deflection, Einstein famously responded ``I would have had to pity our dear Lord.  The theory is correct all the same.''

\subsection{The general-education curriculum}

My college's current curriculum is one typical of a liberal arts college, requiring students to study across all of our academic divisions, engage and improve skills such as written and oral communication, and to draw connections between their courses in different departments.
During our curriculum review process, we have identified that we are in a charmed position in that this curriculum is working well within the limitations of the college's resources.
Thus, a large part of our proposed overhaul has been a reworking of the language that describes our curriculum, so that students and faculty can better understand the importance and goals of its requirements, and to prompt students to exercise more agency in their course choices.
This refining of language has led to some structural improvements, 
but the overall structure of the curriculum will not change much with this new iteration.

If there has been a guiding principle in assembling our new curriculum, it has been that students should be exposed to all of the methods by which humans gather and interpret information about our world.
Others who have gone through similar processes will likely commiserate that defining such general education categories is difficult.
One wants to make a description of, {\em e.g.}, The Arts or The Social Sciences simultaneously powerful and comprehensive, but also accessible to students at any point in their studies (as well as prospective students and parents, if possible).
How does one describe to an eighteen-year-old what The Social Sciences are?  
What knowledge or experiences do teenagers have that can leveraged in this task?
Such descriptions should be brief and digestible, but should also present a strong argument for why the discipline is essential to responsible and engaged citizenship.
I am proud that my colleagues have succeeded in balancing these features and in assembling a curriculum whose primary goal is to guide (rather than just to force) students through our offerings and help them get the most out of their time at our school.
Though I anticipated it to be protracted and painful, our assembly of a description of the Natural Sciences was (relatively) quick and benefited greatly from input from faculty in all departments.

\subsection{Theory: A division in The Natural Sciences}

But a devil was lurking in the details.  
One of the features of our new curricular descriptions is that we seek to give the student some idea of the tools with which knowledge is gathered and evaluated in these fields.
For most of the review process, the following description has been attached to The Natural Sciences:
\begin{quote}
These courses teach students to investigate the principles on which an understanding of the natural world rests, and present the means by which these principles are assembled using the tools of experimental observation, theoretical investigation, modeling, and data collection and analysis.
\end{quote}
This list is not perfect!  
I can't say that I think a student would know what each of these terms means before taking one of my courses, but I do think that a student who successfully completed an introductory physics course could say whether or not she engaged each of these tools.

Toward the end of the process, a colleague who is shepherding the curriculum review contacted me to ask whether dropping ``theoretical investigation'' from the list would present a problem from the perspective of the Physics Department.
The justification given by the review committee for omitting this item was that ``if something is a scientific theory it no longer requires investigation.''
I had to read that phrase a few times to extract its essence.  
This was not the opinion of the person who contacted me; she was trying to paraphrase the (adamant) suggestions of another committee member and several science faculty who had been consulted.

It is rare that one is asked to speak on behalf of one's entire field.  
I am very aware of the reputation that the physics community has for arrogance, a dismissiveness of other scholarly endeavors, an aversion to fool-suffering (in addition to a rather broad definition of ``fool'').
I like to think that I strive to counter this in my professional life.
Though I was quite alarmed by this claim, I proceeded with caution and tried to extend the benefit of doubt.

I assumed that the disconnect here was one of superficial differences in language between fields.
I carefully wrote a response that said that I think theoretical investigation (though ``investigation'' might not be the best word) has been, and still is, an essential part of the accrual of scientific knowledge.
I mentioned that theoretical frameworks give us a way to interpret and understand experimental results, but also help to distinguish between productive and unproductive lines of experimental inquiry.
From theories, we make quantitative predictions, and these predictions are tested by experiment.
I cited examples, such as the development of General Relativity (GR), where theoretical inquiry presaged a revolution in scientific thinking, but needed to be tested by experiment before gaining wide acceptance.
I claimed that when I teach a physics lecture course, these types of developments and the tools that accompany them are what I am teaching.  
I try to be careful to point out the stages of development in a scientific theory, and the experimental evidence that suggested them.

My reply was sent back to the group for discussion.  
Members of our Chemistry and Biology Departments were contacted about whether they would have reservations about dropping any mention of theory from the list.
When I was told that no one voiced any reservations about this, I lost my composure and replied that I was ``baffled.''
My bafflement was met with the following explanation, which I'll reproduce in full for fear of paraphrasing inaccurately:
\begin{quote}
Theories in biology are developed AFTER experimentation and not before. A theory (in the biological sciences) is a fact and is NOT a mere prediction. A concept that has withstood the test of time and repeated experimentation using multiple approaches is accepted as a theory. For example, evolution by natural selection is a theory. Experiments are performed assuming evolution occurs. It is a guiding principle in other investigations, but it in itself is not being investigated. What we investigate are hypotheses and not theories.
\end{quote}
When I read this now, I realize that it was provided partly to clear up my confusion over differences in language.
Physicists, at least in my field, rarely use the term ``hypothesis.'' 
I understand that part of the above claim is that there is a hierarchy of such ideas, and that Biology distinguishes between ``theories'' and ``hypotheses.''

My disagreement with the above claim is an epistemological one.
When I plan and present a course, I have in my mind old-fashioned Popperian notions of theory and evidence\cite{popper}, {\em viz.} that all theories are provisional.
Theories cannot be {\em proven} true by experimentation.
Rather, the best we can do is make predictions based on a theory and determine whether these predictions are {\em consistent} with observation.  
If so, then the theory survives until the next experiment.
Thus, I see my task as being two-find: present the theory and amass an amount of evidence for it such that the theory becomes convincing and (at best) intuitive to the student.
Though quantum indeterminism and more recent shifts in scientific thinking have suggested that falsifiability is not the absolute as presented by Popper, I think that this is the more careful way to educate young scientists. 
It certainly prevents one from claiming that theories are facts.

In physics, theories are constructs that we {\em work} with.
They are vital constructs that guide experimental inquiry, but they {\em are also} tools for inquiry.
Neither experimental or theoretical investigation survives on its own (though there is a vocal post-empiricism faction of theory community).
Theories are approximations of reality; some are quite good, some only work well for specific regimes, and some may stand the test of time and, as a chemist colleague put it, asymptotically approach fact.
The proposed treatment of theories as ``facts'', even if used sloppily in a debate between colleagues, really worries me.

Evolution by natural selection is a particularly nuanced example.
Natural selection is an amazing idea!
It provides an explanation for ({\em i.e.}, is consistent with) many observations that we make of nature, and we {\em see} its effects unfolding in phenomena such as (unfortunately) the evolution of bacteria that are harmful to humans and other species \cite{enright}.
It is a rich way of understanding our world.
However, it is very different from many of the theories that we engage in physics.
Physical theories {\em must} be predictive in a quantitative sense.
Given the framework of GR, one can the dynamics of a binary star system and the gravitational radiation that it emits, which could then be compared to experiment.
Evolutionary theory does not offer the same handholds.
Can we anticipate something like a maximum evolutionary phase for bacteria such as MRSA?
Though it may be possible to develop models that speak to such questions, these models likely aren't contingent upon the veracity of Natural Selection, and thus these models do not grant falsification power.

In physics, theories are testable {\em because} of their quantitative power.
In my opinion, this is an essential lesson in physics pedagogy.
In fact, one of the biggest problems that we have with students in introductory physics classes is helping them to understand that in the lab, they are testing predictions.
These predictions are not exactly sophisticated, and they have already been tested by generations of students before them, but it's important for them to realize that they are doing something much more than just making measurements to characterize a single system.
The theories that they test are applicable to many systems other than the one that they study in a particular experiment, and we must be careful that they do not lose sight of this.

I see another thorny difference between an paradigm such as Natural Selection and many physical theories.
Evolutionary theory is a culturally and politically charged concept.
I think it is epistemologically accurate to say that ``Diversity of species due to evolution by natural selection is a theory; it has not been proven,'' in the sense that a theory cannot be proven by an experiment.
However, this statement carries grave connotations, and has implications for issues such as education and public policy.
There are some circles, both academic and non-academic, in which saying the above would garner the ``Ummm, who invited this guy?'' look.
The same is not true of many similar statements regarding physical theories.
I can not imagine a group that would meet ``Quantum mechanics is a theory; it has not been proven'' with the same disgust.

A chemist at my college has offered a more careful way of thinking about such paradigms.
He suggests that ``evolution is such a powerful explanatory paradigm, the odds are high that it is in fact true.''
I like this statement, though I would wager that there are statisticians that would take offense at how one assigns odds to such things.
I guess you can't please everybody!

\subsection{Theory in introductory physics pedagogy}

At some point during the discussion, a colleague from another science department suggested that the list of tools that we really want is one that describes what our students will actually be doing in their courses (not necessarily those that professionals use), and that an undergraduate curriculum either doesn't have the time or depth to undertake theoretical investigation.
I am sympathetic to this argument.
Biology students need an enormous amount of background knowledge to be able to talk intelligently about their field and be proficient enough to complete advanced courses.
I understand that in order to facilitate this proficiency, classes in biology and introductory chemistry need to be taught in a certain way.
My colleagues' arguments seemed to imply that this is the way that {\em all} science pedagogy works, and that I should adopt a more realistic perspective of my role as a professor.

As I see it, the pedagogical approach is almost completely different in physics.
The introductory physics courses that I teach are predominantly peopled by bright and tenacious biology, chemistry, and psychology students, many of whom are planning a career in medicine.
I often joke with these students that they're in for a treat: they're only going to learn ten or so things in the first semester of physics.  
They can forget about reading the book multiple times, deploying a rainbow fleet of highlighters, or carrying a stack of flashcards.
What they will be expected to do is solve problems -- {\em many} problems.  
They will have to find a way to turn the knowledge that they get from the book and lectures into understanding; this is done by solving problems, looking for patterns, and taking a forest-rather-than-trees approach.
The problems that will be on the exams will {\em not} be the same as the ones that they did in the homework; if they want to succeed, they will need to understand the course material {\em and} be creative with it.
I am deliberate in telling my students these things because this idea of creative application is a huge shift in learning styles for them.

To those that doubt that meaningful theoretical inquiry happens at the introductory level, I offer the following.
Before the mid-1800's, humans did not know what atoms were.  
There were certainly guesses about their existence (dating back to the ancient Greeks!), but empirical evidence was little.
One of the more puzzling features of Nature was gases -- they could be felt, but not seen, and it was clear that they had some relation to other forms of matter due to processes like boiling.  
At the time, many bulk characteristics of gases were known.
When experimenters compressed a gas, they found that its temperature increased, and so forth, according to simple mathematical relationships.
In their second semester of physics, we propose to students an insight: What if gases are made up of tiny particles, too small to see, and they just bounce around inside of a volume and don't interact with one another?
The students, willing to humor their instructor, find that forty minutes later, after applying a little algebra and a few (three) basic relationships from the first semester, they have {\em derived} the empirical gas laws.
This is a compelling suggestion (not proof!) that atoms exist.
In addition, we lead them through assessing what assumptions they've made along the way and when their derivations should be valid.
This is a powerful way of gaining knowledge, and I don't know a way to describe it other than ``theoretical inquiry.''
For students who are able to place themselves in the shoes of those that first worked through such calculations, this exercise is an emphatic lesson about the importance of theoretical inquiry.

\subsection{Theory's place in general education}

Where does this leave the general-education curriculum?
I am thankful that faculty more intelligent and more eloquent than I were consulted on this question, and that ``theoretical inquiry'' has made the list!
But, I'm left feeling a little odd about ``Natural Sciences'' as a general education bucket.
Let me approach the situation this way: I have two hypotheses.

My first hypothesis is obviously silly, but I will present it in the interest of being rigorous, and also because I worry that some of my colleagues think that {\em I think} it is true.
\begin{quote}
{\bf Hypothesis 1:} By virtue of the fact that I am a physicist, my opinions on science and science pedagogy are the correct ones, and I thus have the authority to tell colleagues in other departments how their jobs should be done.
\end{quote}
When I say that this hypothesis is silly, what I mean is that it is not consistent with observations.
The frequency and magnitude with which I've been wrong in the past suggest that this hypothesis should be abandoned.
In addition, the other Natural Sciences departments are also adequately preparing their students for graduate programs and careers, so this hypothesis doesn't present a clear path forward.  
I don't know enough about the courses in the other science departments or about their respective fields; rather, I'm relying on what my colleagues in those departments tell me.
My intuition is that this hypothesis has little to do with reality.

\begin{quote}
{\bf Hypothesis 2:} The learning of physics is something fundamentally different from the learning of biology or chemistry, and we need to be aware of this when speaking of the Natural Sciences as a unified discipline.
\end{quote}
To a degree, I am ambivalent about this statement.
Some courses in my college's chemistry major, such as Physical Chemistry or Quantum Chemistry, are taught in ways very similar to those of our upper-level physics courses: the development of mathematically expressed theories and their experimental evidence are presented.
But, from what my colleagues are suggesting, the modes of learning that students engage in biology and introductory chemistry courses are wildly different from those that they engage in introductory (and later) physics courses.
I doubt that this is a controversial idea, but it does have a potentially striking consequence.
We assume that the goal of curricular requirements is to make our students engage all of the ways by which humans gain knowledge, so that they can act as responsible citizens and appreciate and evaluate information on their own.
If this is so, then the lumping of science courses into one curricular category, requiring students to take only one or two courses from among the Natural Science disciplines, presents an incomplete curriculum.\footnote{It is worth noting that it was also difficult to convince some colleagues that Mathematics is a discipline separate from the Natural Sciences in that mathematical constructs need not be contingent on physical phenomena.}

The story of Einstein and the development of GR is, of course, well beyond the scope of an introductory course for general-education students.
However, theoretical developments such as this, especially those that dramatically change the way that we think about the Universe and our study of it, are powerful ways of knowing and are integral to the progress of science.
Thus, when we teach science, we should be clear about the power of theory.

Our physics department has neither the resources nor the desire to accommodate all of the students at our college.  
Introductory Physics is not a course that's appropriate for every student, mostly (in my opinion) due to the mathematical familiarity it requires.
In addition, I think that this further separation of the natural sciences isn't a productive line of discussion for our current curriculum review.
In my mind, a curriculum is only as good as the academic advising that students receive, and I think that our faculty are very good at this.
I offer these comments simply because they surprised me.
Now that I know that these perspectives exist in higher education, I'll seek ways to make my teaching of theoretical manipulations and philosophy of science more deliberate.

I don't suggest that the anti-theory perspectives described above are shared by the majority of non-physics educators.
But I do take away from this experience a renewed appreciation for the theoretical inquiry that is an essential component of physics.
It is fashionable in physics for experimentalists to grumble about their theorist counterparts and {\em vice versa}, but I think we should be more proud and vocal of this special relationship.
It seems that the relation of the theoretical community in the life sciences has been overlooked at my institution; I don't know if any of our faculty in these departments could be described as ``theorists''.  
For a similar turn to happen in physics would be against the nature of the discipline, and I have a new-found respect for what a great relationship this is.

Science denial is all around us.
Its manifestations range from tacit and annoying (think of the latest fad diet) to bold and catastrophically dangerous (think about the debates over immunization or global warming, or perhaps just the fact that there {\em are} debates about such things).
I don't think that being careless with epistemological concerns helps non-scientists to understand what science allows us to know, and I actually worry that it makes non-scientists more skeptical.
When we're not truthful or up-front about science's limitations, we're not doing an honest job of educating people about its place in our lives.
As science faculty, we often disdain our colleagues in other fields for not being transparent about the scope of their inquiry, and I think that it's time we turn the lens on ourselves.

There are certain assumptions that it seems we must make in order to do productive scientific work.
Perhaps the most fundamental of these is that there exists an objective physical reality for us to describe!
On top of this we stack all manner of other assumptions: causality, locality, {\em et c.}
I would wager that there are very few NSF grant proposals that begin with something like ``Assuming that there is an objective physical reality, and that causality and locality are safe bets, we propose to study the synthesis of insulin in...''
I don't know if I would go to this level in my courses, either.

Perhaps the assumptions made in the Life Sciences about the fundamental nature of principles such as Natural Selection are of the same class.
It seems, however, that physics offers ways of testing the principles that were thought to be basic assumptions.
Indeed, the jury has been out on how to think about locality (the idea that an object can only be influenced by its immediate surroundings) since the mid-1960's; experimentalists and theorists continue to investigate.
To me, the power to test the big ideas is the allure of doing science, and I have seen my students recognize the same.
In addition to the epistemological problem stated above, I think that presenting scientific paradigms as fact gives students the impression that Science is complete.
I can't speak for the Life Sciences, but this is certainly not the case in physics; there are many open questions in the field and some promising experimental and theoretical leads.
Being careful in the classroom gives us a better chance of having more generation of scientists equipped to answer these questions, and a public that supports and values this work.

\vfill
\bibliographystyle{apsrev4-1}
\bibliography{note}

\end{document}